\documentclass[sigconf]{acmart}

\usepackage{graphicx}
\usepackage[linesnumbered, ruled]{algorithm2e}
\usepackage{multirow}
\usepackage{makecell}
\usepackage{enumitem}
\usepackage{amsmath}
\usepackage{subfigure}
\usepackage{algorithmic}
\usepackage{tabularx}
\usepackage{makecell}
\usepackage{hyperref}

\renewcommand\footnotetextcopyrightpermission[1]{} 

\AtBeginDocument{%
  }

\setcopyright{acmlicensed}
\copyrightyear{2018}
\acmYear{2018}
\acmDOI{XXXXXXX.XXXXXXX}

\acmConference[Conference acronym 'XX]{Make sure to enter the correct
  conference title from your rights confirmation emai}{June 03--05,
  2018}{Woodstock, NY}

\begin{document}

\title{Self-supervised Graph Neural Network for \\ Mechanical CAD Retrieval}


\author{Yuhan Quan}
\email{quanyh15@tsinghua.org.cn}
\affiliation{
  \institution{4Paradigm Inc.}
  \city{Beijing}
  \country{China}
}

\author{Huan Zhao}
\email{zhaohuan@4paradigm.com}
\affiliation{
  \institution{4Paradigm Inc.}
  \city{Beijing}
  \country{China}
}

\author{Jinfeng Yi}
\email{yijinfeng@4paradigm.com}
\affiliation{
  \institution{4Paradigm Inc.}
  \city{Beijing}
  \country{China}
}

\author{Yuqiang Chen}
\email{chenyuqiang@4paradigm.com}
\affiliation{
  \institution{4Paradigm Inc.}
  \city{Beijing}
  \country{China}
}

\begin{abstract}

CAD (Computer-Aided Design) plays a crucial role in mechanical industry, where large numbers of similar-shaped CAD parts are often created. Efficiently reusing these parts is key to reducing design and production costs for enterprises. Retrieval systems are vital for achieving CAD reuse, but the complex shapes of CAD models are difficult to accurately describe using text or keywords, making traditional retrieval methods ineffective. While existing representation learning approaches have been developed for CAD, manually labeling similar samples in these methods is expensive. Additionally, CAD models' unique parameterized data structure presents challenges for applying existing 3D shape representation learning techniques directly. In this work, we propose GC-CAD, a self-supervised contrastive graph neural network-based method for mechanical CAD retrieval that directly models parameterized CAD raw files. GC-CAD consists of two key modules: structure-aware representation learning and constrastive graph learning framework. The method leverages graph neural networks to extract both geometric and topological information from CAD models, generating feature representations. We then introduce a simple yet effective contrastive graph learning framework approach, enabling the model to train without manual labels and generate retrieval-ready representations. Experimental results on four datasets including human evaluation demonstrate that the proposed method achieves significant accuracy improvements and up to 100 times efficiency improvement over the baseline methods.

\end{abstract}


\maketitle

\section{Introduction}

Computer-Aided Design (CAD) systems, like Catia\footnote{https://www.3ds.com/products/catia} and Solidworks\footnote{https://www.solidworks.com/}, are instrumental in industrial design and manufacturing. Employed throughout the entire product lifecycle, from conceptualization to production, engineers and designers use CAD software to create accurate 3D models of industrial products and parts. Over time, industrial enterprises can accumulate millions of these CAD designs, also known as parts. However, this vast collection can lead to significant increases in operational costs, including production, testing, and storage. The cost incurred during the entire lifecycle of each new component can vary dramatically, ranging from tens of thousands to hundreds of millions, even billions of euros~\footnote{https://www.cadenas.de/tl\_files/cadenas/Downloads/PDF/Produktflyer/EN\\/CADENAS\_PARTsolutions\_Brochure\_EN.pdf}. Therefore, avoiding the creation of new parts and actively seeking CAD part reuse is crucial for efficient manufacturing practices.

The key to part reuse lies in the CAD retrieval system~\cite{li2010retrieving}. This system assists designers and sales representatives in proactively finding similar parts within a company's database, potentially eliminating the need to create new ones. In a real-world example, the Airbus A380 design team utilized an effective geometric similarity search system. This resulted in a nearly 40\% increase in part reuse rate~\footnote{www.cadenas.de/brochure/geosearch}, savings hundreds of millions of euros. While building a simple retrieval system using text or keywords might seem straightforward, such methods fall short for CAD applications. Unlike searching for documents with search engines like Google, where text accurately conveys content, the complex shapes in CAD models are difficult to capture fully with words. An effective CAD retrieval system requires the ability to search based on 3D shape similarity, posing a significant challenge in practice.

In the literature, various works have been proposed to address the aforementioned problem. One common strategy involves designing feature descriptors based on the 3D geometry of the model. There are many variations in the design of descriptors, such as multi-view based on different viewpoints~\cite{wang2019dominant, qi2016volumetric}, moments~\cite{elad2002content,kazhdan2002harmonic} or spherical harmonics~\cite{vranic2001tools,saupe20013d}, etc. These methods all use the global features of the 3D shape, and some methods try to model based on the topological information of the shape, such as modeling shape as graphs~\cite{bespalov2003reeb,el2003database} or trees~\cite{bai2010design}. After extracting features by descriptors, histograms, graphs or trees describing CAD shape features can be obtained. Then, the similarity score between CAD parts can be obtained by calculating the similarity based on these descriptors. However, such heuristic methods are highly dependent on domains and human expertise; thus, they are neither cheaper nor generalized in practice.
Moreover, for some complex descriptors, such as graphs, graph matching algorithms are generally NP-hard \cite{hilaga2001topology}, which means that the similarity calculation will be very time-consuming.

Recent advances in deep learning have shown promise for various CAD tasks, including segmentation~\cite{lambourne2021brepnet,jones2023self} and assembly~\cite{willis2022joinable}, due to their improved generalization capabilities and computational efficiency. To adopt deep learning, most works translate CAD parts into graph structures~\cite{jones2023self, willis2022joinable, mandelli2022cad}. However, there is no learning method specifically designed for CAD similarity retrieval. Furthermore, implementing deep learning for CAD retrieval faces a significant hurdle in data labeling. Compared to text documents, labeling CAD parts for similarity necessitates deeper expertise in mechanical engineering, considerably increasing the cost.\footnote{We ourselves encountered this challenge: ensuring high-quality labels required recruiting at least bachelor-level mechanical engineering students.} Scaling this approach to train deep learning models for CAD retrieval is impractical, especially considering the vast number of parts (potentially millions) often present in large manufacturing companies.

In general, there are currently two challenges in building an effective CAD similarity retrieval system:

\begin{itemize}
 \item \textbf{Limited feature extraction}: Existing methods, whether text-based or relying on 3D data descriptors, often fail to fully capture the rich geometric and topological information inherent in CAD models. This leads to poor generalization capabilities and can be computationally expensive.
  
 \item \textbf{High labeling cost}: Despite the good generalization ability of learning-based methods, it remains a challenging problem to train an effective one, considering that the labeling cost of CAD similarity labels is extremely high. Therefore, an unsupervised learning method is crucial to make CAD retrieval practical and scalable.
  
\end{itemize}

Addressing the critical challenges of capturing both geometric and topological features in CAD retrieval, and the high cost of manual labeling, we introduce GC-CAD (Graph Contrastive learning for CAD retrieval) in this work. This novel self-supervised graph learning framework models CAD parts directly from their BRep format as graphs (see Figure \ref{fig:brep_intro} and Section \ref{subsec-preli}), leveraging graph neural networks for representation learning. By employing an effective contrastive learning method, GC-CAD utilizes unlabeled CAD data, avoiding expensive labeling. Extensive experiments on public and private datasets, including real-world applications, demonstrate the superiority of GC-CAD.

To summarize, the contributions are as follows:

 \noindent$\bullet$ \textbf{First self-supervised GNN for CAD retrieval}: To the best of our knowledge, we are the first to address the CAD similarity retrieval problem using self-supervised GNN, which can improve the retrieval capability while avoiding the necessity of labeling works. Moreover, it is also an important and novel problem in terms of GNN research.  

  \noindent$\bullet$ \textbf{Effective and label-free approach}: We develop an effective graph contrastive learning framework, including a structure-aware representation learning module and contrastive training pipeline, which enables the model to learn from the geometric and topological information of CAD shapes without the need for any labels.

  \noindent$\bullet$ \textbf{Extensive evaluation}: We rigorously evaluate GC-CAD on four diverse datasets comprising up to 300,000 CAD shapes. Our experiments, including human evaluation, demonstrate that our method outperforms existing baseline approaches in terms of both accuracy and efficiency.

\section{Related works}

\subsection{CAD Retrieval} 

To develop effective CAD retrieval systems, the key challenge is to model 3D shapes of a CAD part with proper descriptors that can be processed or indexed. As a unique data format, CAD files store parameterized shape information, such as point coordinates, curve equations, surface equations, etc. CAD shape can be described from both global and local aspects. For global features, multi-views~\cite{wang2019dominant, qi2016volumetric}, moments~\cite{elad2002content,kazhdan2002harmonic} or spherical harmonics~\cite{vranic2001tools,saupe20013d} can be used. These methods usually rely on statistical characteristics of 3D shapes, which are not accurate when the shape is complex. In order to describe the local geometric features and the overall topology of CAD, a common method is to model CAD as a graph or tree. ReebGraph is a classical solution to deal with CAD parts directly~\cite{bespalov2003reeb, hilaga2001topology}. The core of ReebGraph is to divide the surface of a 3D shape into small units and construct the graph according to the adjacency relationship. In addition, graphs or trees can also be constructed based on the adjacency relationship~\cite{chen2012flexible} or hierarchical relationship~\cite{bai2010design} of cad local features. In recent years, with the development of graph neural networks, some works have tried to directly extract the graph structure that can be used by graph neural networks from the topology structure of CAD. The common idea is to represent the faces in CAD as nodes on the graph, and the adjacency relationship between faces (i.e., the topology information of CAD) is represented as edges on the graph~\cite{jayaraman2021uv, mandelli2022cad}. Furthermore, some elements such as CAD vertices and loops can also be used to construct more complex graphs~\cite{jones2021automate, cao2020graph}.

After the representation of a CAD part is obtained, there are two classes of methods for similarity retrieval of CAD: heuristic and embedding-based methods. For heuristic methods, the core is to compare the descriptors extracted from CAD for retrieval. Commonly used techniques include histogram features~\cite{ankerst19993d, berchtold1998section} and graph features~\cite{bespalov2003reeb, hilaga2001topology, di2002new}. Among them, histogram is often used to represent statistical features, which is naturally not accurate enough and ignores the local features of CAD. While comparing whether two graphs are similar is actually a graph isomorphism problem, and it is NP-hard to obtain an accurate solution~\cite{babai2016graph}, which means that the computational cost may be very high. Embedding-based methods solve these problems. After using graph neural networks to learn the embeddings~\cite{jayaraman2021uv, lambourne2021brepnet, jones2021automate, colligan2022hierarchical}, similarity can be measured by distance functions or other scoring methods. However, in order to learn the representations, deep neural networks often need a large amount of annotated data for training, and obtaining a large amount of annotated data in CAD field is highly expensive. Therefore, the proposed GC-CAD framework makes the first attempt to address these challenges facing CAD retrieval by developing a graph contrastive learning framework, which is totally free of any labeling works and is highly generalized due to the adoption of GNN.

Note that beyond BRep, there are other lines of parallel works modeling 3D shapes in other formats, like voxel and meshes\cite{maturana2015voxnet, xu2016beam,feng2019meshnet}, point cloud \cite{qi2017pointnet,qi2017pointnet++,klokov2017escape}, multi-view image \cite{wang2019dominant, qi2016volumetric, kanezaki2018rotationnet} (see more in \cite{gezawa2020review}) .
However, these methods are not suitable for CAD-related tasks in mechanical domains because of the information loss \cite{lambourne2021brepnet} by converting BRep to the corresponding format. Moreover, it is more convenient to integrate a method that directly on BRep topology with existing working pipelines in manufacturing enterprises since BRep is more common in CAD and PLM \footnote{https://www.ptc.com/en/products/windchill-plus}(Product Lifecycle Management) systems. Therefore, from a technical perspective, these works are orthogonal and complementary, and the position and contribution of the proposed GC-CAD can be clearly distinguished.

\subsection{Self-supervised Graph Neural Networks}

Graph neural networks play an important role in many fields such as social networks~\cite{chen2022gdsrec, quan2023robust}, chemical molecules~\cite{wang2023automated}, and knowledge bases~\cite{ji2021survey}. Due to the huge amount of data (such as social networks) or the need for domain knowledge (such as chemical molecules), it is not easy to obtain a large number of labels in all fields, which gives rise to the need for unsupervised graph neural networks. Unsupervised learning can be divided into two categories: contrastive and generative~\cite{xie2022self}. Contrastive learning is a widely applied self-supervised method~\cite{chen2020simple,he2020momentum}. In graph learning, data augmentation includes modifying graph topology structure at the graph level~\cite{zeng2021contrastive, suresh2021adversarial} or modifying node features at the node level~\cite{perozzi2014deepwalk,zhu2020deep}. Then a loss function based on mutual information maximization can be constructed. For generative methods, the loss function is constructed by reconstructing the input data and features, similarly, including node features~\cite{hu2019strategies, you2020does, hou2022graphmae} or graph structure~\cite{hasanzadeh2019semi, pan2018adversarially}. For CAD retrieval problem, since each CAD part is usually modeled as a graph, the computation is actually at the graph level. However, in addition to the overall topology information, the local features of CAD also need to be taken into account, which means that node features are also needed. Both graph-level and node-level objectives need to be considered in designing self-supervised learning tasks. 

In this paper, to the best of our knowledge, we are the first to introduce self-supervised GNN into an important and new real-world applications, i.e., CAD part retrieval,  by addressing two critical challenges, which also push forward the frontiers of GNN research.
Note that there is indeed a recent work~\cite{jones2023self} trying to learn the representation of CAD through unsupervised learning. However, it is not specifically designed for similarity retrieval, and it uses signed distance field(SDF) as the target of generative learning and optimizes by regression loss, which is not suitable for capturing shape similarities between CAD parts. In the experiments, the performance advantage of the proposed GC-CAD demonstrates the superiority of contrastive training manner.

\section{Method}

\begin{figure*}
	\centering
	\includegraphics[width=.88\textwidth]{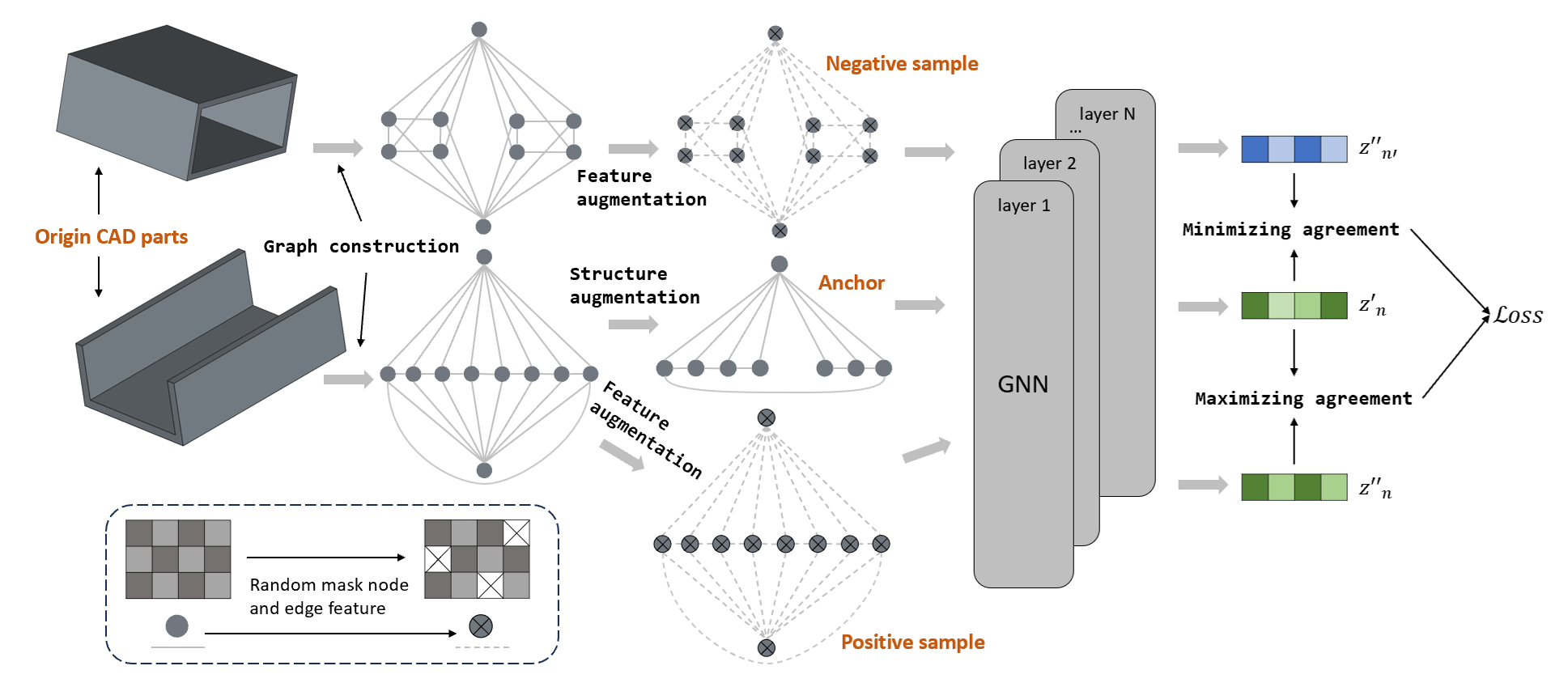}
	\caption{Overall framework of GC-CAD. Each CAD part is first converted to a graph, then feature and structure augmentation are performed separately. Augmented graph is input to GNN to obtain graph representation. Finally, GC-CAD constructs and learns from positive and negative sample pairs based on whether the graph representation are from the same original CAD part. After the training is completed, the CAD retrieval can be performed using graph representation.}
	\label{fig:framework}
	\vspace{-10pt}
\end{figure*}

In this section, we introduce GC-CAD, which mainly includes structure-aware representation learning and self-supervised training. We start from the BRep of CAD part, convert it to the graph, and then extract features from it through GNN, so as to obtain representation containing geometric and topological information of CAD parts. Then, we design a self-supervised contrastive learning method. The goal of contrastive learning is to enable the model to distinguish the representation from different CAD parts. Finally, the trained GNN model is used to generate meaningful representation of CAD part, and the similarity retrieval is realized by the vector retrieval method. The overall framework is shown in Figure~\ref{fig:framework}.

\subsection{Notation}

Let $P=\{p_1, p_2, ... \}$ denote a set of CAD parts, where $p_i$ denotes a part. For the CAD similarity retrieval problem, the goal is to retrieve all similar parts $\{p_r| p_r\in P \& p_r \neq p_q \}$ from $P$ based on a given query $p_q$, where $p_r$ is similar with $p_q$ in shape. We denote $g_i=\{V,E\}$ as the graph extracted from the CAD part $p_i$, where $V$ and $E$ are the sets of nodes and edges, respectively. $\mathbf{x^v}$ and $\mathbf{x^e}$ denotes the original feature of graph nodes and edges, while $\mathbf{\hat{x}}^v$ and $\mathbf{\hat{x}}^e$ denote the dense vector obtained by processing original features with neural networks. $\mathbf{h}$, $\mathbf{e}$ and $\mathbf{z}$ denote the node embedding, edge embedding and graph embedding of graphs respectively.

\begin{figure}[t]
	\centering
	\includegraphics[width=.44\textwidth]{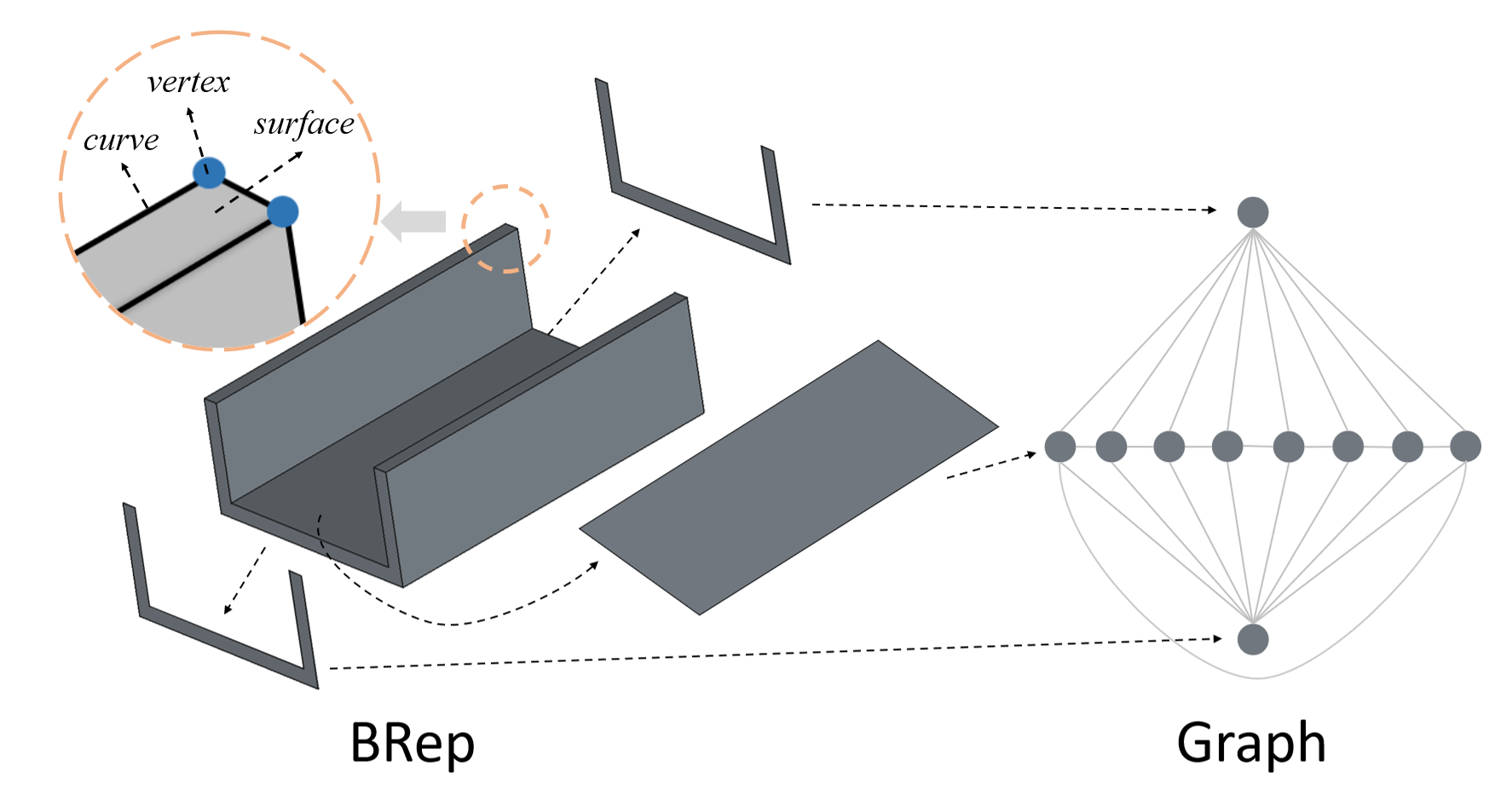}
	\caption{Boundary representation(BRep) and converting it to graph. The faces and curves in BRep correspond to the nodes and edges in graph.}
	\label{fig:brep_intro}
	\vspace{-15pt}
\end{figure}

\subsection{Preliminaries}
\label{subsec-preli}

Boundary representation(BRep) is a common representation method in CAD. It represents a solid by a set of faces, curves and vertices, where faces intersect to form curves, and curves intersect to form vertices. Boundary representation describes the features of these geometric objects and the topological relations between them, and these information together define the shape of a CAD entity. CAD files store the parameters that define these geometric objects and the relationships between them. Naturally, topological relations can be described by graph structures, while the features of each geometric object itself are represented as features on the graph. In line with existing work~\cite{jayaraman2021uv}, each face on the CAD shape is represented as a node on the graph, and each curve is represented as an edge on the graph. Figure~\ref{fig:brep_intro} shows an example. Note that we are assuming that every curve has two and only two faces adjacent to it, which means that every edge can connect exactly two nodes, and there are no hyperedges in graph. This assumption is true in most cases, but for more complex cases, we can either ignore such curves or split one curve into multiple \footnote{In some CAD systems, for conical or cylindrical surfaces, there may exist a bus line, which is adjacent to only one surface, and such a curve need not be embodied in the graph. Whereas if a curve is adjacent to more than two surfaces, we build an edge between the nodes corresponding to each of the two faces.}.

\subsection{Structure-aware representation Learning}

\subsubsection{Feature extraction}
After converting the topological structure of a CAD part into a graph, we need to extract the geometric information to describe the local features of the part. In this step, we refer to UVNet~\cite{jayaraman2021uv} as backbone. It extract geometric information includes uv-grids features $\{\mathbf{x}_{uv}^e,\mathbf{x}_{uv}^v\}$ and parameterized geometric features $\{\mathbf{x}^{v}_{geo}, \mathbf{x}^{e}_{geo}\}$. We do not repeat the details of these features. In addition to these features used by uvnet, except the shape, the material, processing method, color and other features of the part also play important roles in CAD retrieval, these product features on each face are represented as $\mathbf{x}^{v}_{product}$.

\subsubsection{Graph neural networks}

After obtaining all the original features, next step is to convert the original features into node embeddings and edge embeddings on graph. We use CNN for uv-grids features $\mathbf{x}_{uv}^v$ and $\mathbf{x}_{uv}^e$. For $\mathbf{x}^{v}_{geo}$, $\mathbf{x}^{e}_{geo}$ and $\mathbf{x}^{v}_{product}$, we transform discrete values into embeddings and normalize continuous values. These features are then input into the MLP to obtain dense embedding. Finally, the dense embedding of different original features are concatenated to obtain the final node embedding and edge embedding. The process is as follows:
\begin{gather}
    \mathbf{\hat{x}}_{uv}^v = CNN(\mathbf{x}_{uv}^v),\  \mathbf{\hat{x}}_{uv}^e = CNN(\mathbf{x}_{uv}^e),  \nonumber \\
    \mathbf{\hat{x}}^{v}_{geo} = MLP_1^v(\mathbf{x}^{v}_{geo}),\  \mathbf{\hat{x}}^{e}_{geo} = MLP_2^e(\mathbf{x}^{e}_{geo}),  \nonumber \\
    \mathbf{\hat{x}}^{v}_{product} = MLP_3(\mathbf{x}^{v}_{product}), \nonumber \\
    \mathbf{\hat{x}}^e = cat([\mathbf{\hat{x}}_{uv}^e, \mathbf{\hat{x}}^{e}_{geo}]), \nonumber \\
    \mathbf{\hat{x}}^v = cat([\mathbf{\hat{x}}_{uv}^v, \mathbf{\hat{x}}^{v}_{geo}, \mathbf{\hat{x}}^{v}_{product}]),\nonumber
\end{gather}

Note that the feature $\mathbf{x}^{v}_{geo}$ and $\mathbf{x}^{e}_{geo}$ differ due to different types of surfaces and curves, as $MLP_1^v$ and $MLP_2^e$ do not share parameters for different types of surfaces and curves, while $CNN_{1D}$, $CNN_{2D}$ and $MLP_3$ share parameters for all surfaces and curves. $CNN$ not only contains convolutional layers but also includes commonly used pooling layers and fully connected layers. We do not describe these commonly used network structures in detail.

After obtaining node embedding and edge embedding containing geometric information, combined with graph structures containing topological information, we use graph neural networks to obtain graph representations $\mathbf{z}$, which represents the overall shape of CAD part. Specifically, taking node embedding $\mathbf{\hat{x}}^u$/$\mathbf{\hat{x}}^v$ and edge embedding $\mathbf{\hat{x}}^e$ as the input of the first layer, denote as $\mathbf{h}^{(0)}_u$/$\mathbf{h}^{(0)}_v$ and $\mathbf{e}^{(0)}_{uv}$, we perform the convolution operation in $K-th$ layer of GNN as follows
\begin{equation}
  \mathbf{h}^{(K)}_v = f^{K}[\mathbf{h}^{(K-1)}_v+sum\{f_\theta(\mathbf{e}^{(K-1)}_{uv}) \cdot \mathbf{h}^{(K-1)}_u, v \in \mathcal{N}(u)\}], \nonumber 
\end{equation}
where $f^{K}$ is an MLP and $\mathcal{N}(u)$ represents the set of all neighbor nodes of node $u$, $f_\theta$ is the activation function. Furthermore, the edge embedding of each layer is updated in the following way
\begin{equation}
  \mathbf{e}^{(K)}_{uv} = g^{K}_1[\mathbf{e}^{(K-1)}_{uv}+g^{K}_2(\mathbf{h}^{(K-1)}_u+\mathbf{h}^{(K-1)}_v)], \nonumber
\end{equation}
where $g^{K}_1$ and $g^{K}_2$ are MLP. Through such a message-passing process, after $K$ layers of convolution, each node can receive the information of nearby nodes and curves so that the embedding of each node contains the local geometric and topological information corresponding to the CAD part. Finally, we use an aggregation operation, based on the node embedding to obtain the final graph representation:
\begin{equation}
  \mathbf{z} = \sum_{v\in V} \sum_{k=1}^K w^{(k)} \cdot \mathbf{h}^{(k)}_v+b^{(k)}, \nonumber
\end{equation}
where $w^{(k)}$ is the parameter. Here, we used the node embedding of each layer in GNN for aggregation, this is because the node embedding of different layers contains the information of local regions with different ranges. Now, for each CAD part, we obtain a representation modeling its geometric and topological information, which can be used for similarity retrieval.

\subsection{Contrastive Training}

\subsubsection{Data augmentation}
The similarity labeling between CAD parts is expensive, which requires experts to judge for each query to traverse all samples in the database. At present, there is no public dataset containing CAD similarity labels. To solve this problem, we design a self-supervised contrastive learning method that enables the model to learn without labels by augmenting the graph, constructing training pairs and then training. For data augmentation, We augment the graph in two aspects: features and structure. For features, for each node and edge on the graph, we randomly mask part of the original features, and for structure, we design three augmentation schemes:
\begin{itemize}
  \item Randomly remove nodes.
  \item Randomly remove nodes and their 1-hop neighbors.
  \item Randomly remove edges and the incident two nodes.
\end{itemize}
Since nodes and edges on a graph correspond to faces and curves in CAD, these data augmentation operations mean the removal of part of the CAD part. We only conduct delete operations without adding operations, because adding random faces or curves in CAD often leads to illogic, and even some faces or curves may penetrate the entity, such CAD part is meaningless. For each graph, after performing feature and structure augmentation separately, two augmented graphs can be obtained whose representations can be denoted as $\mathbf{z}'$ and $\mathbf{z}''$. 

\subsubsection{Model training}
Furthermore, in model training, $2N$ augmented graphs can be obtained for each mini-batch of size $N$. Based on this, using the NT-Xent loss function, for the $n-th$ graph in a mini-batch, its optimization objective is as follows~\cite{you2020graph}:
\begin{gather}
  l_n=-\log \frac{\exp (\operatorname{sim}(\mathbf{z}'_{n}, \mathbf{z}''_{n}) / \tau)}{\sum_{n^{\prime}=1, n^{\prime} \neq n}^N \exp (\operatorname{sim}(\mathbf{z}'_{n}, \mathbf{z}''_{n'}) / \tau)},  \nonumber \\
  \operatorname{sim}(\mathbf{z}'_{n}, \mathbf{z}''_{n})=\frac{(\mathbf{z}'_{n})^{\top} \mathbf{z}''_{n}}{\|\mathbf{z}'_{n}\|\|\mathbf{z}''_{n}\|}, \nonumber
\end{gather}
where $\tau$ denotes the temperature parameter. This loss function maximizes the agreement between representations of two data-augmented graphs from the same original CAD part, while the agreement between representations that do not originate from the same original CAD part is minimized. By optimizing the loss function, we can make the model learn from geometric and topological information to distinguish graph representation from different CAD parts, so as to obtain representation that can be used for retrieval. 

\subsubsection{Inference and application}
In application, given a query CAD part, similarity retrieval is implemented by computing the similarity (e.g., cosine similarity, l2 distance, etc.) between its representation and the representations of other samples in the database. At the same time, faster retrieval can also be achieved by vector retrieval tools such as faiss~\cite{johnson2019billion}. Compared with heuristic methods, this greatly improves the efficiency.

\subsection{Discussion}

We adopted contrastive over generative learning for our self-supervised task, leveraging data augmentation on node, edge, and graph structure to differentiate representations without specific prediction targets. Generative approaches were bypassed due to their complexity in balancing local and global feature learning—essential for CAD retrieval. Our method simplifies training by masking node and edge features, efficiently capturing both local and global CAD features for contrastive learning. In the experiments, the superiority of contrastive over generative framework is verified in Table \ref{tab:mian_results}.

Furthermore, for the data augmentation of contrastive learning, we add mask to the node and edge features and graph structure. Although this operation seems simple, for CAD parts, these operations correspond to modifying the local shape and global structure of the CAD part while still making the augmented CAD part as meaningful as possible, which is suitable for contrastive learning. In the experiments, the superiority of contrastive over generative framework is verified in Table \ref{tab:mian_results}.

\begin{table*}[t]
	\centering
	\caption{Overall performance of GC-CAD compared with baselines on two private datasets.}
	\begin{tabular}{c|ccccc|ccccc}
		\toprule
		Dataset & \multicolumn{5}{c|}{Dataset A} & \multicolumn{5}{c}{Dataset B} \\
		Metric & Recall@5 & Recall@10 & NDCG@5 & NDCG@10 & Time & Recall@5 & Recall@10 & NDCG@5 & NDCG@10 & Time \\
		\midrule
		ReebGraph & \textbf{52.1\%} & 56.5\% & 63.1\% & 53.9\% & \(\sim\) 7d & 38.4\% & 39.0\% & 56.2\% & 47.1\% & \(\sim\) 14d \\
		BRepNet & 40.8\% & 42.8\% & 52.7\% & 44.1\% & \(\sim\) 1.5h & 35.2\% & 36.0\% & 51.0\% & 42.8\% & \(\sim\) 2.5h\\
		SS-CAD & 49.6\% & 53.8\% & 62.8\% & 52.7\% & \(\sim\) 3h & 38.2\% & 39.1\% & 56.4\% & 46.8\% & \(\sim\) 5h \\
		\midrule
		GC-CAD & 51.5\% & \textbf{56.8\%} & \textbf{65.0\%} & \textbf{55.9\%} & \(\sim\) 2h & \textbf{39.3\%} & \textbf{39.8\%} & \textbf{58.0\%} & \textbf{48.7\%} & \(\sim\) 3h \\
		\bottomrule
	\end{tabular}
	\label{tab:mian_results}
\end{table*}
\section{Experiment}

\subsection{Baselines}
To the best of our knowledge, we are the first work to address the CAD retrieval problem in a self-supervised GNN framework. Thus there are not too many baselines taht can be directly compared. To verify the effectiveness of the proposed GC-CAD in label efficiency and generalization, we then choose a total of three baseline methods, including heuristic methods, kernel methods and self-supervised learning methods.

 \noindent$\bullet$ \textbf{ReebGraph}~\cite{bespalov2003reeb} is a heuristic method, we convert each sample into a ReebGraph, and then for each query, use graph matching algorithm to align it with each candidate sample and compute the similarity score, thus achieving retrieval.
  
  \noindent$\bullet$ \textbf{BRepNet}~\cite{lambourne2021brepnet} defines some kernels based on BRep, then extracts features from them by convolution, and finally obtains the representation of surfaces and curves. This is a model designed for the segmentation task, we pool the representations of surfaces and curves output by BRepNet to obtain a representation that can represent the entire CAD, making this method useful for CAD retrieval. At the same time, due to the lack of training labels, we combine this method with our proposed self-supervised method to be able to train the model.
  
  \noindent$\bullet$ \textbf{SS-CAD}~\cite{jones2023self} is a generative unsupervised method for learning a representation of a CAD, which constructs an optimization objective by sampling points near the surface of the CAD and predicting the signed distance  field(SDF) function value for this point. The sign and magnitude of the SDF function values can indicate whether a point is located inside or outside a closed solid, as well as the distance of the point from the surface of the entity. Therefore, this method can make the model learn the CAD shape information.

\subsection{Experiment settings}
\textbf{Datasets}: We used a total of four datasets for experiments, including two public datasets and two private datasets. Two private datasets are derived from our internal sources and named Dataset A and Dataset B respectively~\footnote{Due to confidentiality requirements, we cannot present sample cases for these two datasets, and we only report the value of metrics for performance comparisons.}. For public datasets, to the best of our knowledge, there is no publicly available dataset for CAD similarity retrieval. We used the ABC dataset~\cite{Koch_2019_CVPR} and JoinABLe Assembly-Joint dataset~\cite{willis2022joinable}. ABC is currently the largest publicly available CAD dataset containing 1 million CAD files without any similarity labels. We filtered this dataset by removing assemblies and keeping only those samples that contained one entity in each file. We also filtered files larger than 1M by file size. For JoinABLe Assembly-Joint dataset, each sample in this dataset contains two parts that are assembled together, and we take these parts apart so that each sample contains only a single part. The statistics of all datasets are shown in the Table~\ref{tab:statistics_dataset}.

\textbf{Evaluation and metrics}
For private datasets, we randomly sampled 40 and 60 samples from the two datasets as queries respectively. For each experience and each query, we retrieve a total of 100 samples as results, and annotate these 100 retrieved results with three labels: similar, partially similar and dissimilar. It is used to measure to what extent the retrieved results are similar to the query sample. All annotators are master's degree candidates in mechanics. The labeling work took about 50 person-days for all experiments in all. Finally, for each sample, we calculate Recall@5, Recall@10, NDCG@5 and NDCG@10 separately and finally report the average value as the result.

For public datasets, we randomly sampled some samples as queries in these two datasets for experiments, but due to the high cost, we did not conduct human evaluation, but to showcase the comparisons with the top 5 retrieved results.

For sake of space, more implementation details are given in Appendix~\ref{appendix:implementation}.

\begin{table}
  \centering
  \caption{Statistics of datasets. The number of faces and curves correspond to the number of nodes and edges in graph.}
  \begin{tabular}{cc|ccc}
  \toprule
  & Dataset & \#num & \#avg faces & \#avg curves \\
  \midrule
  \multirow{2}*{Private} & Dataset A & 6,060 & 72.4 & 355.5 \\
                         & Dataset B & 8,922 & 83.4 & 409.9 \\
  \midrule
  \multirow{2}*{Public}  & ABC       & 360,424 & 26.6 & 122.6 \\
                         & JoinABLe  & 22,913  & 36.5 & 171.4 \\
  \bottomrule
  \end{tabular}
  \label{tab:statistics_dataset}
  \vspace{-10pt}
\end{table}

\begin{table*}[t]
	\centering
	\caption{Results of ablation study for different structure augmentation method.}
	\vspace{-5pt}
	\begin{tabular}{c|cccc|cccc}
		\toprule
		Dataset & \multicolumn{4}{c|}{Dataset A} & \multicolumn{4}{c}{Dataset B} \\
		Metric & Recall@5 & Recall@10 & NDCG@5 & NDCG@10 & Recall@5 & Recall@10 & NDCG@5 & NDCG@10 \\
		\midrule
		Node                 & \textbf{51.5\%} & \textbf{56.8\%} & \textbf{65.0\%} & \textbf{55.9\%} & \textbf{39.3\%} & 39.8\% & \textbf{58.0\%} & 48.7\%  \\
		Node\&1-hop neighbor & 50.1\% & 56.0\% & 63.4\% & 54.8\% & 38.6\% & 39.7\% & 57.4\% & 48.7\% \\
		Edge\&Vertices       & 50.6\% & \textbf{56.8\%} & 63.4\% & 54.3\% & 39.0\% & \textbf{41.6\%} & 57.7\% & \textbf{49.1\%} \\
		\bottomrule
	\end{tabular}
	\label{tab:ablation_mask}
\end{table*}

\subsection{Performance Comparison}

The experimental results based on human evaluation on two private datasets are shown in Table~\ref{tab:mian_results}. We trained and tested the model on both datasets separately. For the proposed GC-CAD, we tested three different data augmentation methods and selected the one that performed best. From the table we emphasize the following three findings:

\noindent$\bullet$	GC-CAD achieves the best results on almost all metrics. Especially on dataset B, compared with the best baseline, the proposed method achieves an improvement of about 1\%. On dataset A, the NDCG metrics are improved by about 2\% compared with Reebgraph. While ReebGraph is close to GC-CAD on Recall metrics, it takes significantly more time than GC-CAD and other deep learning-based methods. On average, it takes more than 12 hours to retrieve each query, because reebgraph needs to compare with all samples in the dataset and calculate the similarity score, where graph matching algorithm is very time-consuming. By contrast, GC-CAD can compute representations and retrieve based on vectors, which achieves nearly 100 times improvement in computational efficiency compared with ReebGraph.
 
 \noindent$\bullet$ BRepNet performs significantly worse than GC-CAD, which demonstrates the superiority of graph-based modeling over other methods. Although BRepnet directly uses parameterized CAD information, it extracts features based on a predetermined kernel. The kernel is designed for segmentation and pays more attention to local features, which is insufficient in modeling the overall topological features compared with GC-CAD and thus performs poorly on similarity search tasks.
 
 \noindent$\bullet$ SS-CAD based on generative self-supervised learning and GNN performs worse than GC-CAD and ReebGraph but better than BRepNet. This shows that graph-based methods are able to capture both geometric and topological information of CAD. However, the optimization goal of SS-CAD is to predict the SDF value of points on CAD, and the SDF value of each point can only represent local features, which makes the model pay more attention to local features and ignore global topology information. In addition, compared with contrastive learning to classify samples, SDF is a continuous value and requires the model to perform regression prediction, which makes the training of the model more difficult, so the performance on the similarity retrieval task is not good.

\begin{figure}[t]
	\centering
	\includegraphics[width=.4\textwidth]{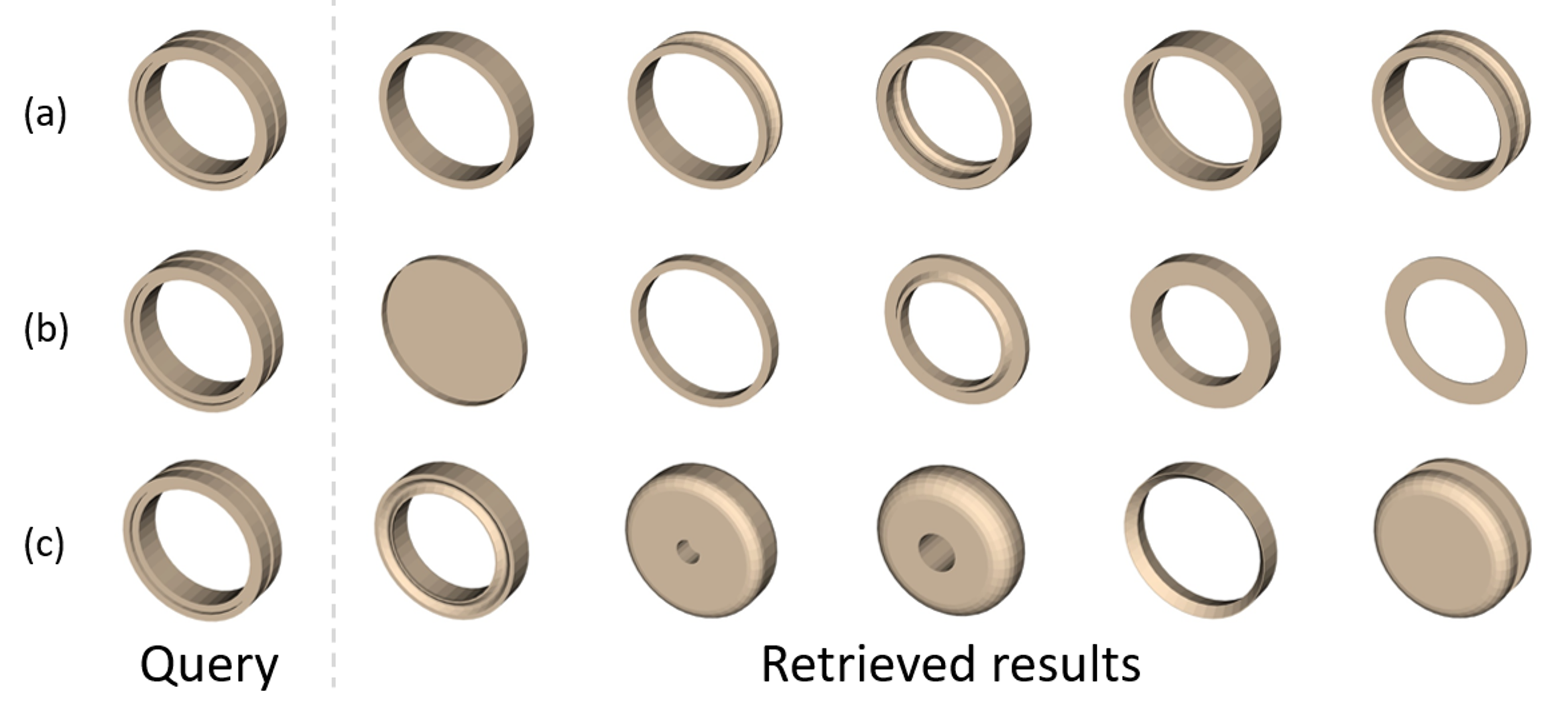}
	\caption{Retrieved results of different data augmentation method.(a) mask node. (b) mask node and its 1-hop neighbors. (c) mask edge and its vertices.}
	\label{fig:mask_ablation}
\end{figure}

\subsection{Ablation and Hyperparameter Study}

As an important module of contrastive learning, we further explore the method and hyperparameters of data augmentation on retrieval results.
\subsubsection{Graph Augmentation Methods}

We first conduct experiments to verify the influence of different data augmentation methods on the retrieved results. We test our proposed three methods of modifying the graph structure, and the experimental results are shown in Table~\ref*{tab:ablation_mask}. It should be noted that for different data augmentation methods, we control the hyperparameters so that the number of nodes removed is consistent across all experiments. The experimental results show that simply random masking node is the best, and random masking node and its first-order nodes are the worst. We think that this is because if a node and its adjacent nodes are masked, it means that some adjacent faces are deleted in CAD, which may make the local features of CAD missing, so that the model cannot correctly distinguish different CAD, while only random masking nodes can keep the local features of CAD as much as possible. We also conducted experiments on the JoinABLe dataset. Figure~\ref*{fig:mask_ablation} shows an example. It can be found that although the retrieved results are all circular, the local features of CAD are not consistent in the results (b) and (c). Therefore, random masking nodes are a simple but effective scheme for CAD retrieval task.

 \begin{table}[t]
	\centering
	\caption{Ablation study of different mask ratio on Dataset A}
	\begin{tabular}{cc|ccc}
		\toprule
		\multicolumn{2}{c|}{\multirow{2}*{\makecell{Recall@5/\\NDCG@5}}}& \multicolumn{3}{c}{$\beta$}  \\
		&  & 0.0 & 0.1 & 0.2 \\
		\midrule
		\multirow{3}*{$\alpha$} & 0.0 & /             & 48.3\%/61.2\% & 48.5\%/61.7\% \\
		& 0.1 & 47.7\%/60.0\% & \textbf{51.5\%}/\textbf{65.0\%} & 48.3\%/61.0\% \\
		& 0.2 & 51.4\%/63.6\% & 53.2\%/64.8\% & 48.8\%/61.7\% \\
		\bottomrule
	\end{tabular}
	\label{tab:ablation_ratio_a}
	
\end{table}

\begin{table}[t]
	\centering
	\caption{Ablation study of different mask ratio on Dataset B}
	\begin{tabular}{cc|ccc}
		\toprule
		\multicolumn{2}{c|}{\multirow{2}*{\makecell{Recall@5/\\NDCG@5}}}& \multicolumn{3}{c}{$\beta$}  \\
		&  & 0.0 & 0.1 & 0.2 \\
		\midrule
		\multirow{3}*{$\alpha$} & 0.0 & /             & 38.2\%/56.2\% & 39.1\%/\textbf{58.3\%} \\
		& 0.1 & 38.5\%/56.2\% & \textbf{39.3\%}/58.0\% & 38.9\%/57.3\% \\
		& 0.2 & 37.5\%/56.0\% & 38.1\%/56.8\% & 38.7\%/57.6\% \\
		\bottomrule
	\end{tabular}
	\label{tab:ablation_ratio_b}
\end{table}

\subsubsection{Feature Mask Ratio}

We further verify the influence on the retrieved results of different ratio of masking features and nodes. We conducted a grid search over the feature masking ratio and node masking ratio, and the experimental results are shown in Table~\ref{tab:ablation_ratio_a} and \ref{tab:ablation_ratio_b}. The experimental results on the two datasets show that it works best when $\alpha$ and $\beta$ are both set to 0.1. Using only one feature augmentation, or using a large ratio of masks, will result in poor retrieval performance. In order to further show the influence of different hyperparameters on the retrieved results, we also conduct experiments on the JoinABLe dataset, and the experimental results are shown in Appendix~\ref{appendix:ablation}. It can be found that for the first column, that is, mask only the features but not the graph nodes, the retrieved results in only local shape(the fold shape) that have similarities to the query, while the overall shape differs greatly. On the other hand, if we mask the graph nodes without the features, i.e., the first row, the retrieved results are cylindrical, but the details differ. The best retrieved results are obtained when both $\alpha$ and $\beta$ are 0.1. We try to increase the hyperparameter value further and the retrieved results become worse. The reason might be that masking too many features and structures will impair the topology of CAD parts, thus the model has been completely unable to distinguish between positive and negative samples. Especially, if these two values are increased (greater than $0.2$), because too many features are masked, the retrieved results will be significantly inconsistent with the query, regardless of the overall shape or local details, resulting in poor performance, thus we only keep the range in $(0,0.2)$ in our experiments.

\subsection{Results on more diverse scenarios}

\begin{figure}
	\centering
	\includegraphics[width=.4\textwidth]{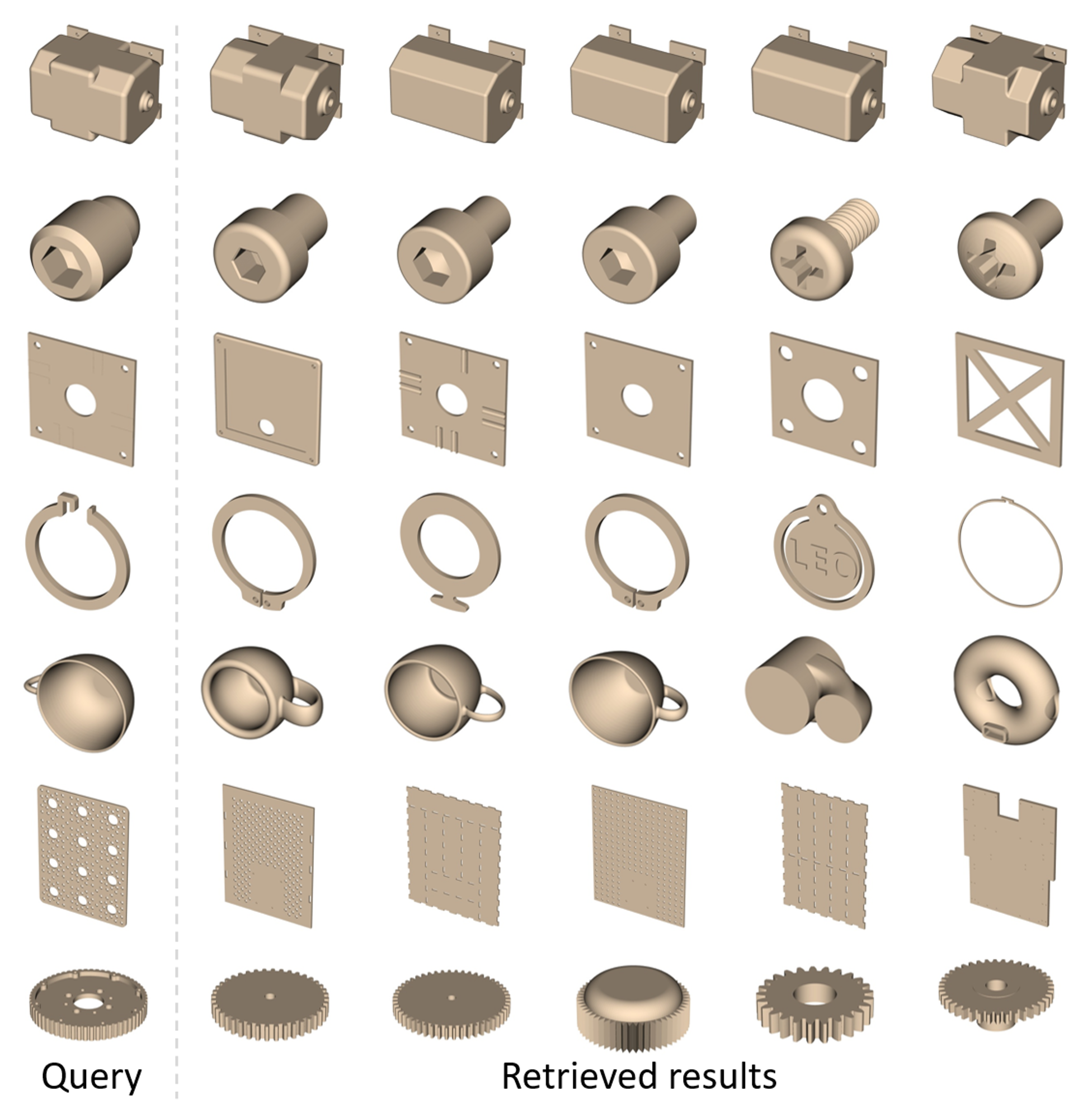}
	\caption{Retrieved results on large scale dataset.}
	\label{fig:abc_res}
	\vspace{-15pt}
\end{figure}

\begin{figure*}
	\centering
	\includegraphics[width=.8\textwidth]{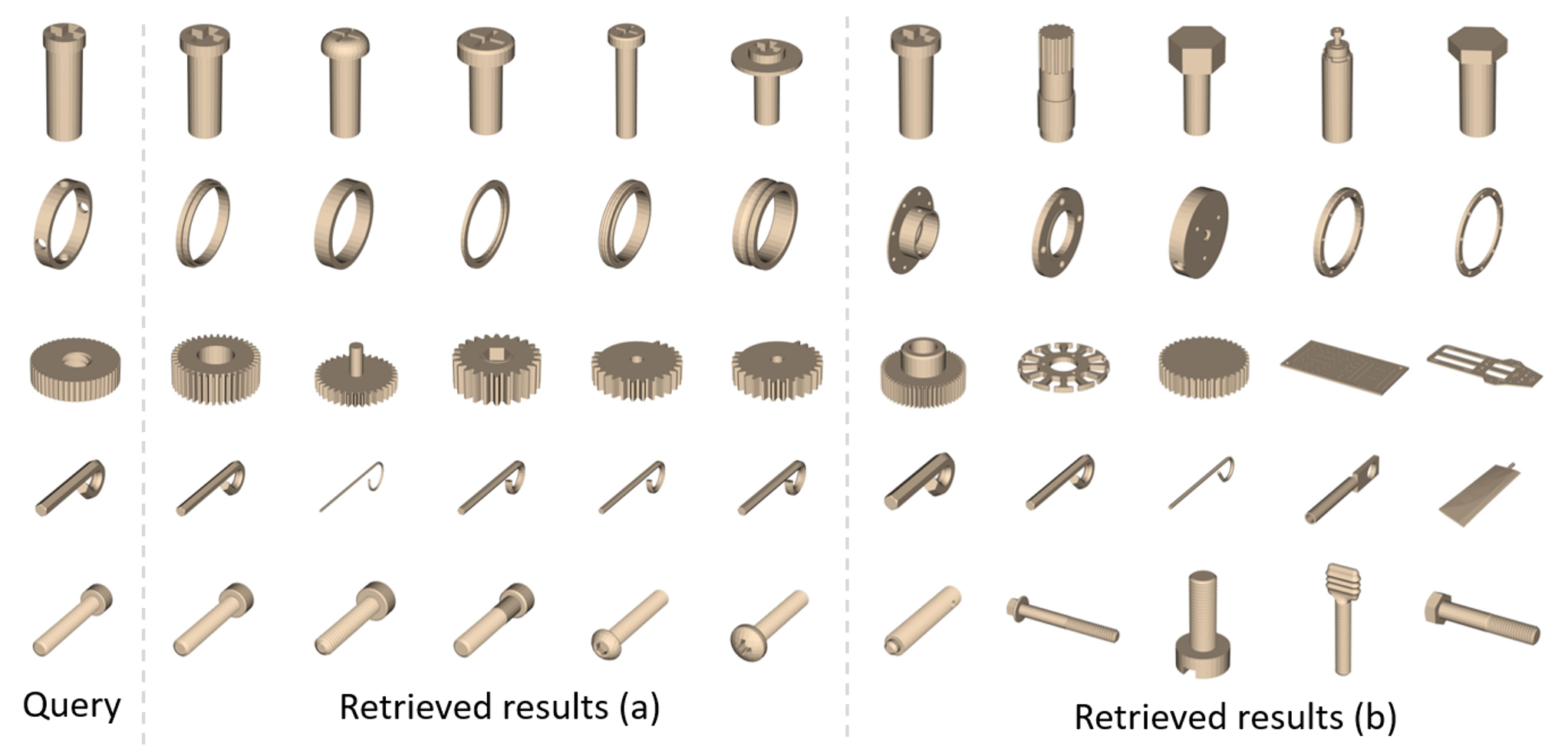}
	\caption{Retrieved results for generalization experiments.(a)Training on ABC dataste. (b)Training on ABC\_small dataset.}
	\label{fig:generalization_res}
		\vspace{-10pt}
\end{figure*}

To further verify the effectiveness of the proposed GC-CAD, we further conduct experiments on various real-world manufacturing scenarios including large-scale number of parts, generation on totally new scenarios, and assembly retrieval.

\subsubsection{Large-scale number of CAD parts.}
To verify the effectiveness of GC-CAD on large-scale datasets in real-world scenarios, we further conduct experiments based on ABC dataset~\cite{Koch_2019_CVPR}, which is currently the largest known public CAD dataset. We train the model with over 360,000 CAD files without using any similaity labels. We randomly sampled some samples as queries, and the experimental results are shown in Figure~\ref{fig:abc_res}. As the figure shown, even on very large datasets and without using any labels, our proposed GC-CAD still has excellent retrieval performance, able to retrieve similar samples from hundreds of thousands of candidate samples based only on the shape of the query. And since the representation of each CAD can be computed in advance, after constructing the index, using a vector retrieval like FAISS~\cite{johnson2019billion}, millisecond retrieval can be achieved, which is completely impossible with heuristic methods like ReebGraph.

\subsubsection{Generalization scenario.}

After verifying the feasibility of GC-CAD on large-scale datasets, inspired by the fact that large language models can be trained on massive datasets to improve generalization performance, we further explore the feasibility of large-scale pre-training GC-CAD to improve generalization performance, so that the pre-trained models can be directly applied to retrieval tasks on new CAD datasets. We construct the ABC\_small dataset ($\sim$30,000 CAD parts) by randomly sampling $10\%$ parts from the ABC dataset, and performing unsupervised training with both datasets separately, keeping the hyperparameters strictly the same. After that, without any fine-tuning, we tested the trained model on the JoinABLe dataset, and the results of the two models are shown in Figure~\ref{fig:generalization_res}. We selected some queries to retrieve using the two models respectively. It can be found that even without using any JoinABLe dataset to participate in the training, the model trained with ABC dataset still shows good performance, and the retrieved results are significantly better than the results obtained by training with ABC\_small dataset. This demonstrates that GC-CAD has better generalization ability with more data in training stages, which also leave an interesting future direction: can we keep increasing the performance by further increasing the number of parts, say from millions to billions?

\begin{figure}
	\centering
	\includegraphics[width=.4\textwidth]{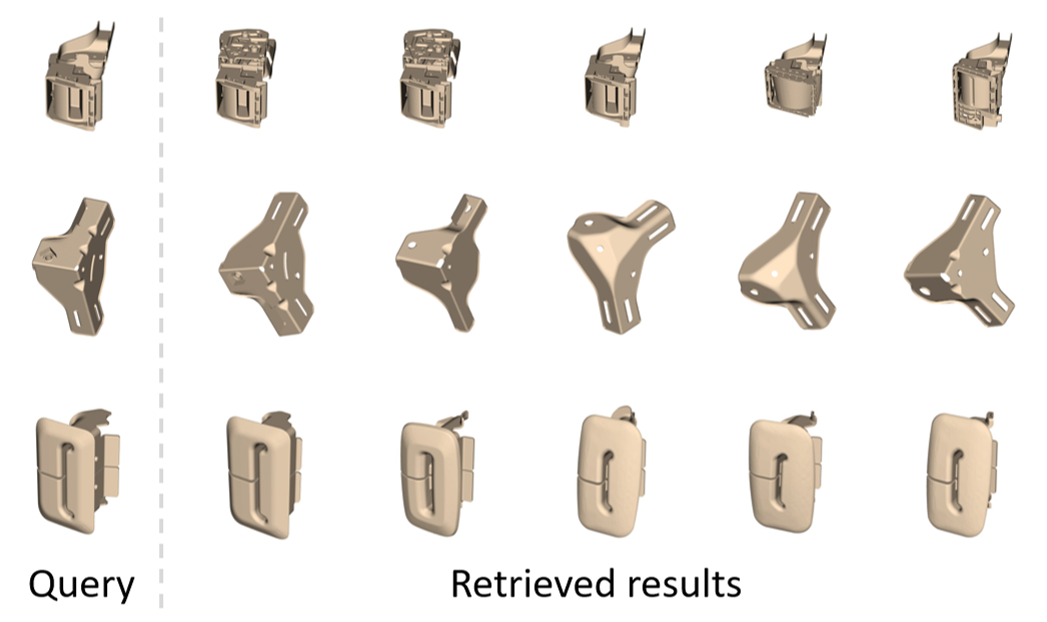}
	\caption{Retrieve results of assemblies. More details of the assembly are shown in the Appendix~\ref{appendix:assembly}.}
	\label{fig:assembly_res}
	\vspace{-20pt}
\end{figure}

\subsubsection{Assembly retrieval beyond single part.}

Here we further explore a bit more complex scenario: \textit{assembly retrieval}. An assembly is formed by multiple parts through assembly, and various parts need to be combined to form a useful module. Assembly retrieval requires attention not only to the shape of each entity, but also to the relationship between entities. There has been some work on assembly retrieval~\cite{deshmukh2008content, lupinetti2019content}. We tested a naive approach that achieves assembly retrieval based on GC-CAD by individually retrieving each part of an assembly and then aggregating and ranking the results. In detail, we disassemble all assemblies into single parts, then for the query assembly, for each of its parts, we retrieve similar entities from the database, and then for each entity in retrieved results, we find in which assemblies it has been used. After performing the retrieval operation on all the parts of the query assembly, we sort the found assemblies according to the number of occurrences. Therefore we realize the retrieval of the assemblies. We use a private dataset mainly of automobile parts to verify the approach. This dataset contains about 500 assemblies and we randomly select some assemblies as queries for retrieval. Instead of training a new model, we still use the model obtained by training on the ABC dataset. The results are shown in Figure~\ref{fig:assembly_res}. It can be seen that our proposed method can effectively retrieve assemblies with similar shapes on the whole or assemblies with similar entities. .
\vspace{-5pt}

\section{Conclusion}

In this paper, we present GC-CAD, a novel method for CAD similarity retrieval that leverages contrastive learning and graph neural networks. Our approach addresses the critical challenge of simultaneously modeling both geometric and topological features for effective retrieval. In addition, we introduce a simple yet effective contrastive learning scheme to eliminate the need for expensive and time-consuming manual labeling, making it highly scalable and practical. Experimental results on both private and public datasets with human evaluation demonstrate the effectiveness of our method, and the computational efficiency is up to 100 times faster than the baseline method. We further validate the effectiveness of GC-CAD on large datasets and assembly retrieval tasks. In the future, we plan to explore data augmentation methods specifically tailored to CAD properties to enhance retrieval accuracy.

\clearpage

\bibliographystyle{ACM-Reference-Format}
\bibliography{ref}

\clearpage

\appendix

\section{Implementation details}
\label{appendix:implementation}
\textbf{Hyper parameters}: For all experiments, the node embedding size and graph embedding size are set to 128 and 256, respectively. The batch size is set to 32 due to the GPU memory limit. For all deep learning methods, we use the Adam optimizer. Hyper-parameters like learning rate and dropout are obtained by grid search. The search range is as follows: the learning rate is \{0.005, 0001, 0.0005\}, the dropout is [0, 1], and the temperature of NT-Xent loss is \{0.5, 1.0, 2.0\}. For all experiments of GC-CAD, we train the model for at least 20 epochs, the early stop is set to 10. The feature mask ratio $\alpha$ and structure mask ratio $\beta$ in data augmentation are all searched in \{0.0, 0.1, 0.2\} and the number of GNN layers is 5. 

\textbf{Running environment}: All experiments are conducted on a Linux server with Intel(R) Xeon(R) Silver 4214 and 512G RAM. We only use a single NVIDIA RTX 3090 GPU for all deep learning methods.

\section{Ablation study}
\label{appendix:ablation}

\begin{figure*}
	\centering
	\includegraphics[width=.88\textwidth]{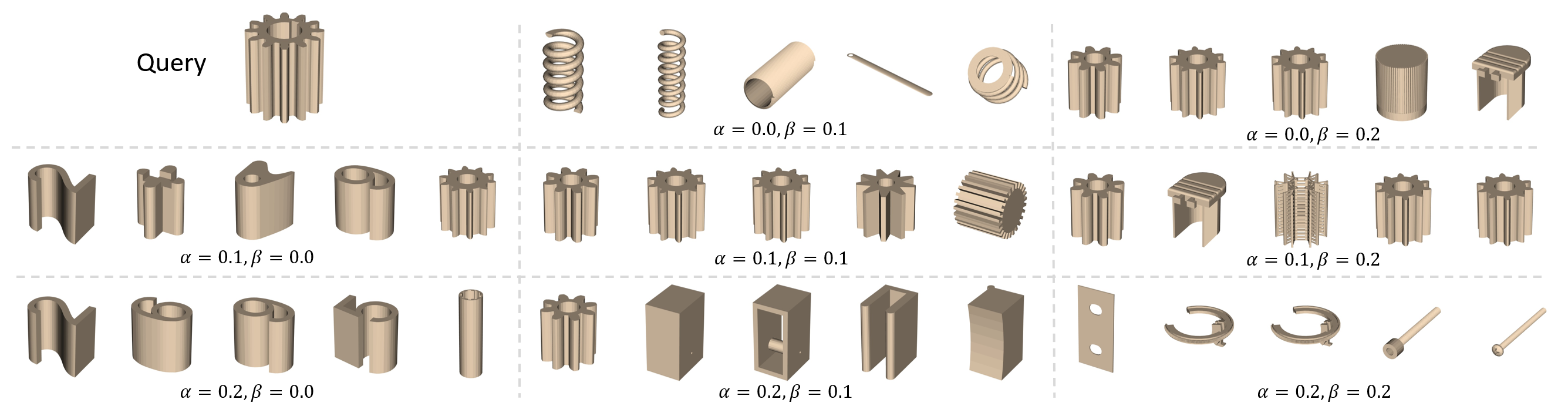}
	\caption{Retrieved results of different mask ratio for data augmentation on JoinABLe dataset.}
	\label{fig:ratio_ablation}
\end{figure*}

Retrieved results of different mask ratio for data augmentation on JoinABLe dataset are shown in Figure~\ref{fig:ratio_ablation}.

\section{Details of assembly}
\label{appendix:assembly}
\begin{figure*}
  \centering
  \includegraphics[width=.6\textwidth]{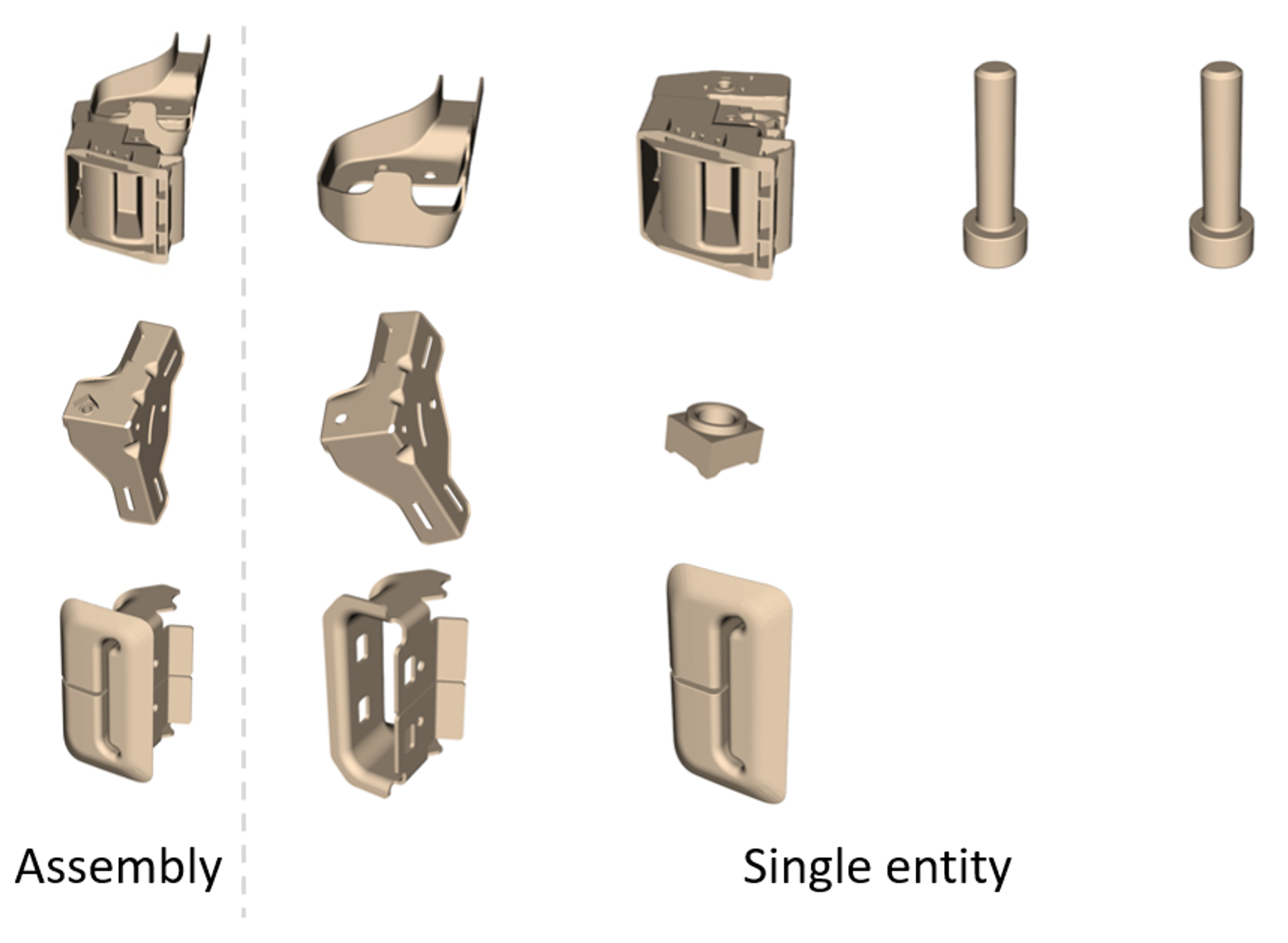}
  \caption{Details of query assemblies.}
  \label{fig:assembly_entity}
\end{figure*}

We disassembled the assemblies in the assembly retrieve experiment to better observe the details, and the results are shown in Figure~\ref{fig:assembly_entity}. It should be noted that we zoom in on smaller entities for ease of observation.

\end{document}